\documentclass[12pt]{article}        %%%%%%%%%%%%%%%%%%
\usepackage{amsmath}
\usepackage{amssymb}

\usepackage{epsfig}

\usepackage{cite}

\usepackage{epic}
\usepackage{fancybox}

\begin{document}

\input{feynman}
\renewcommand{\theequation}{\arabic{equation}}
\def\fnote#1#2{\begingroup\def\thefootnote{#1}\footnote{#2}\addtocounter
{footnote}{-1}\endgroup}
\topmargin=-.5truein \textheight=8.6in \oddsidemargin=-.25in
\evensidemargin=-.25in \textwidth=6.7in
\def\txs{\textstyle} \def\half{{\txs{1\over2}}}
\def\np#1{{\sl Nucl.~Phys.~\bf B#1}}\def\pl#1{{\sl Phys.~Lett.~\bf #1B}}
\def\pr#1{{\sl Phys.~Rev.~\bf D#1}}\def\prl#1{{\sl Phys.~Rev. Lett.~\bf #1}}
\def\cpc#1{{\sl Comp.~Phys. Comm.~\bf #1}} \def\be{\begin{equation}}
\def\anp#1{{\sl Ann.~Phys.~(NY) \bf #1}} \def\ee{\end{equation}\ }
\def\etal{{\it et al.}}\def\sqstev{$\sqrt s=40$~TeV}
\def\bbar#1{\llap{\phantom#1}^{\scriptscriptstyle(}\bar#1^{\scriptscriptstyle)}}
\def\Kmax{K_{\rm max}}\def\ieps{{i\epsilon}}\def\rQCD{{\rm QCD}}
\def\rngp{renormalization group\ }
%\begin{document}
\hfill UTHEP--97--1001\vskip.01truein
\hfill {Oct., 1997}\vskip0.5truein
\centerline{\Large Size of Penguin Pollution of the CKM CP Violating}
\centerline{\Large Phase in
     $\bar B_s\rightarrow \rho K_S$   \fnote{$\ast$}{Research supported in part
by the US DoE ,grant     DE-FG05-91ER40627 and contract DE-AC03-76SF00515.
          }}\vskip0.18truein
\centerline{\sc B.~F.~L.~Ward  }
\vskip.13truein
\centerline{\it Department of Physics and Astronomy}
\centerline{\it The University of Tennessee, Knoxville, TN 37996--1200}
\centerline{\it and }
\centerline{\it SLAC, Stanford University, Stanford, CA 94309}
\centerline{\it USA}
\vskip.05truein \baselineskip=25pt\vskip0.25truein
\centerline{\bf ABSTRACT}\vskip.1truein\par
We use the perturbative QCD methods of Lepage and Brodsky
to calculate
the rate for $\bar B_s\rightarrow \rho K_S  $, with an eye toward the CP
violating unitarity triangle angle  $\gamma$. We show that , although
the penguins are large, there are regions of the allowed
parameter space of the Cabibbo-Kobayashi-Maskawa mixing matrix wherein
$\gamma$ is measurable 
in the sense that penguins change the value
of $\sin(2\gamma)$ one would extract from the attendant time dependent
asymmetry measurement by less than $29\%$, so that a 3$\sigma$ measurement
of $\sin(2\gamma)$ as being different from $0$ is allowed by the
corresponding theoretical uncertainty. This
would establish CP violation in $B_s$ decays.
The rates which we find tend to favour
the type of luminosities now envisioned for hadron-based B-factories.
 
\par\renewcommand\thepage{}\vfill\eject
\parskip.1truein \parindent=20pt \pagenumbering{arabic}
%\section{Introduction}
\par
Now that there are two asymmetric $e^+e^-$ colliding beam B-Factories,
the SLAC-LBL-LLNL and KEK Asymmetric B-Factories, as well as several
other B-Factory type machines, such as HERA-B, the CESR upgrade,
and the Tevatron upgrade, for example, under 
construction, the systematic exploitation of these machines
for  CP violation studies is not far away. 
To realize the true potential of these studies, it is
important that the complete set of
Standard Model CP violation parameters for the B-system be explored,
if it is at all possible. In particular, this means that all CP violating
angles $\alpha, \beta$ and $\gamma$ of the unitarity triangle should be
measured ,where we use the notation of ~\cite{sbfacwkp} for these angles.
The angle $\beta$ is the "gold plated" angle of the triangle, as
it will be presumably the most readily measurable of the three
angles, via the modes $B\rightarrow \Psi/J K_S,\Psi/J K^*_+$. It ($\beta$)
is in fact used to specify the minimal requirements
for the B-Factory machine and detector system to be
successful. (Here, ${K^*}_+$ denotes the CP + neutral $K^*$
meson.)
Accordingly, the B decay modes needed for measurement of the angles $\alpha$ and
$\gamma$ must also be identified and assessed. In this connection,
the mode 
$\bar B^0_s\rightarrow \rho+ K_S  $ is worthy of some
attention; for, were it not for the possible contamination from penguins,
this mode would be a candidate mode for 
the measurement of $\gamma$~\cite{sbfacwkp}. 
Indeed, the potential contamination from penguins is just as substantial as it is for the mode $\bar B\rightarrow \pi^0\pi^0$ in connection 
with the measurement
of $\alpha$, for which the authors~\cite{pentrap}
have devised isospin methods to combine the measurements of
the   modes $B\rightarrow \pi^+ \pi^-,\pi^0 \pi^0$ and
$B^+\rightarrow \pi^+ \pi^0$ to extract $\alpha$ independent of the
size of the penguin contamination-- the main experimental problem of
course is the measurement of the $\pi^0 \pi^0$ mode.
It is desirable to address these penguin CP violation pollution
effects from a dynamical approach which aims to quantify them directly,
thereby isolating just where a measurement may still be made, in view
of the available parameter space in the respective
Cabibbo-Kobayashi-Maskawa mixing matrix. Indeed, in a recent paper~
\cite{pipi}, we analysed the theoretical expectations for the size of
these penguins in the basic mode $\bar B\rightarrow \pi^+ \pi^-$
as well as in the companion mode $\bar B\rightarrow \pi^0 \pi^0$.
We have found that, in a large region of the parameter space,
the Asymmetric B-Factory devices at SLAC and KEK will be able
to extract the fundamental
CP violating angle $\alpha$ without depending on the penguin trapping
methods in Ref.~\cite{pentrap}. The natural question to ask is
whether an analogous region exists in the case of the measurement
of the angle $\gamma$ in the $B_s\rightarrow \rho K_S$ decay? It is
this question that we address in the following theoretical development.
\par
 
\indent
Specifically, we will use the approach of Lepage and Brodsky~\cite{L-B}, 
as it is represented in our analysis of $D\rightarrow  \pi^+ \pi^-$,
$K^+ K^-$ in Ref.~\cite{dppkk}. In this realization of perturbative
QCD for hard exclusive processes, as we shall illustrate explicitly
below, the exclusive amplitude is represented as a convolution
of a hard scattering kernel (referred to as $T_H$
in Ref.~\cite{L-B}) with distribution amplitudes that
sum the respective large QCD collinear logarithms associated
with radiation from the external legs of the constituent partons.
These distribution amplitudes therefore obey a rigorous QCD evolution
equation derived from QCD perturbation theory in Ref.~\cite{L-B}.
We refer to this representation of hard exclusive hadron processes
as the Lepage-Brodsky method. It was already formulated
in Ref.~\cite{shbrod} in the context of the exclusive two-body
B decays to light mesons of the type of interest to us here.
See also Refs.~\cite{simwylr,ccarlsn} for further illustrations
of the method we shall use.
As we explain in Ref.~\cite{pipi},
we expect the accuracy of our methods as used here to be at least
as accurate as the $25\%$ accuracy determined in the work in 
Ref.~\cite{dppkk}.
We present both the absolute decay rates and the ratio 
of branching ratios corresponding to 
such rates, with and without the penguins included in the
respective calculations. In this way, we expect to minimise the sensitivity of
of our results to the uncertainty of the normalisation of
the distribution amplitudes which we do use. Indeed, in the
respective  CP asymmetry
parameter $\sin(2\gamma)$ analysis, we compute its apparent shift
away from its expected value in the absence of penguins in
ratio to that expected value, $\Delta\sin(2\gamma)/\sin(2\gamma)$.
We refer to this shift as the penguin shift of $\sin(2\gamma)$.
The analog of this shift plus unity was already introduced 
in Ref.~\cite{gronau} in the 
study of the time-dependent CP-violating asymmetry in the $\pi^+\pi^-$
decay mode. We will exhibit
a formula for the penguin shift of $\sin(2\gamma)$ here 
for definiteness in complete analogy with what
we have already published in Ref.~\cite{pipi} for the corresponding
shift of the analogous CP violating asymmetry parameter $\sin(2\alpha)$ 
for the $\pi^+\pi^-$ decay mode.
Evidently, the normalisation of our distribution amplitudes also drops out
of the penguin shift of $\sin(2\gamma)$.
\par
\indent
Concerning the Cabibbo-Kobayashi-Maskawa (CKM) matrix itself,
we follow the conventions of Gilman and Kleinknecht 
in Ref.~\cite{G-K} for the
CP-violating phase $\delta_{13} \equiv \delta$ and in view of the
current limits on it we consider the entire range 
$0\leq \delta \leq 2\pi$.  For the CKM matrix parameters $V_{td}$ and
$V_{ub}$ we also consider their extremal values from Ref.~\cite{G-K}
(the Particle Data Group (PDG) compilation).
To parametrise these
extremes, we use the notation defined in Ref.~\cite{rfleisch}
for $|V_{ub}/V_{cb}|$
in terms of the parameter $R_b= .385\pm .166$~\cite{G-K}.
All other CKM matrix 
element parameters are taken at their central values~\cite{G-K}. 
\par 
\indent
We should emphasise that the decay under study here is
not the only way to study the CP-violating angle $\gamma$.
Indeed, due to the very small rates which we shall find, 
it will be seen that the most appropriate machine to pursue
the mode under discussion here is a hadron collider type
B-factory device. As shown in Refs.~\cite{sbfacwkp,rfleisch,manfleis}, the
$e^+e^-$ colliding beam type B-factory device can approach
$\gamma$ from other decay mode avenues.
\par
\indent
We further emphasise that it is possible to use the methods of 
Lepage and Brodsky~\cite{L-B} ,as they
are represented in the analyses in Refs.~\cite{pipi,dppkk}, to address both 
the concept of colour suppression
for the $\bar B_s\rightarrow \rho K_S $ decay as well as the size of
the penguin pollution in its CP violating phase structure as 
described above. We will take advantage of this opportunity
to get a quantitative estimate of the colour suppression effect
in this decay under study here. In practice, what this will
mean is that, in addition to computing our branching ratio (BR) for the
decay with and without penguins included, we will also
compute it with and without gluon exchange
between the would-be spectator $\bar s$ and the $q\bar q$ lines of
the outgoing $\rho$. Again, we will focus on the respective ratios
of BR's to avoid sensitivity to the uncertainty in the normalisation
of our distribution amplitudes.
Such an estimate of colour suppression has not
appeared elsewhere.
\par
\indent
Specifically, we note that the QCD corrections to the weak 
interaction Lagrangian will be represented via
%be implemented after the ideas of Bauer,Stech and Wirbel (BSW) in 
%Ref.~\cite{BSW},
%whereby the field current identity is used to interpolate the
%$\rho$ into the operator $O_2=Q_1$ in 
the QCD corrected effective
weak interaction Hamiltonian ${\cal H}_{eff}$
as it is defined in Ref.~\cite{rfleisch}
\begin{equation}
{\cal H}_{eff} = {G_F\over\sqrt{2}}\left[\sum_{j=u,c}{V^*}_{jq}V_{jb}
                  \left\{\sum^{2}_{k=1}Q^{jq}_k\tilde C_k(\mu)+
                   \sum^{10}_{k=3}Q^{q}_k\tilde C_k(\mu)\right\}\right]+h.c.
\end{equation}
where the Wilson coefficients $\tilde C_i$ and operators $Q_k$ are as given
in Ref.~\cite{rfleisch}, $G_F$ is Fermi's constant, $\mu$ is 
is the renormalization scale and is of ${\cal O}(m_b)$ and here $q=s$.
The application of this effective weak interaction Hamiltonian
to our process $\bar B_s\rightarrow \rho K_S$ then proceeds according to
the realization of the Lepage-Brodsky expansion as
described in Ref.~\cite{shbrod}. This leads to the ``dominant''
contribution in which the $\rho$ is interpolated into the 
operator $O_2=Q_1$ in ${\cal H}_{eff}$ via the factorised
current matrix element $<\rho|\bar u(0)\gamma_\mu P_L u(0)|0>$,
~$P_L\equiv\frac{1}{2}(1-\gamma_5)$
so that the
respective remaining current in $O_2=Q_1$ is responsible for the $\bar B_s$ to
$K_S$ transition shown in Fig.1.  We refer to this contribution as
the ``Tree'' contribution.
The complete amplitude for the
%%%%%%%%%%%%%%%%%%%%%%%%%%insert fig.1%%%%%%%%%%%%%%%%%%%%%%%
\begin{figure}
\begin{center}
\epsfig{file=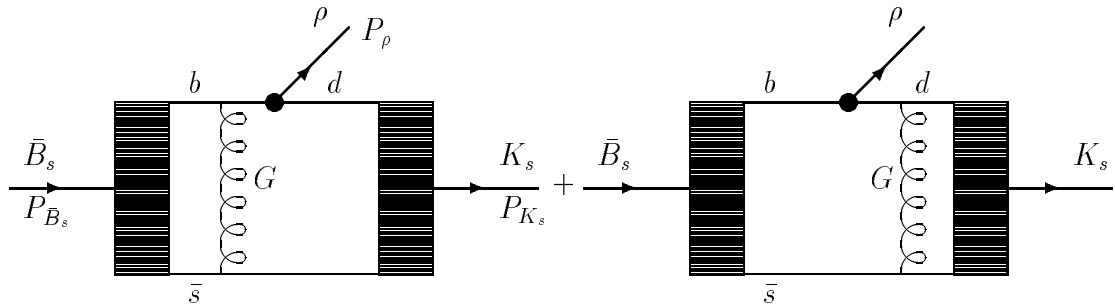}
\end{center}
\vspace{ -1.00cm}
\caption{\baselineskip=7mm     The process $\bar B_s
           \rightarrow           \rho
                 +       K_S$.
The four--momenta are indicated in the standard manner: $P_A$ is the
four--momentum of $A$ for all $A$.To leading order in the
perturbative QCD expansion defined by Lepage and Brodsky in
Ref.~\cite{L-B}, the two graphs shown are
the only ones that contribute in the dominant contribution
as isolated by the methods of Ref.~\cite{shbrod}
when penguins and colour exchange between the outgoing $\rho$
partons and the outgoing $K_S$ partons are ignored. The
remaining graphs in which the gluon $G$ is exchanged between
the would-be spectator $\bar s$ and the remaining $\rho$ parton lines
as well as the penguin type graphs are shown in Figs. 2 and 3,
where we see that, for QCD penguins, there is the added possibility 
that the gluon $G$ interacts with the penguin gluon itself of course.
}
\label{figone}
\end{figure}
%-----------------------------------------------------------------------------
%-----------------------------------------------------------------------------
%-----------------------------------------------------------------------------
%-----------------------------------------------------------------------------
%-----------------------------------------------------------------------------
%%%%%%%%%%%%%%%%end figure 1%%%%%%%%%%%%%%%%%%%%%%%%%%%%%%%%%%%% 
\noindent
process under study here,
$\bar B_s\rightarrow \rho K_S$, is given by the sum of the
graphs in Fig.~\ref{figone} and those in 
Figs.~\ref{figtwo} and \ref{figthree},
to leading order in the Lepage-Brodsky expansion defined in Ref.~\cite{L-B}
and realized according to the prescription in Ref.~\cite{shbrod}.
%----------------begin fig 2--------------------------------
\begin{figure}
\begin{center}
\epsfig{file=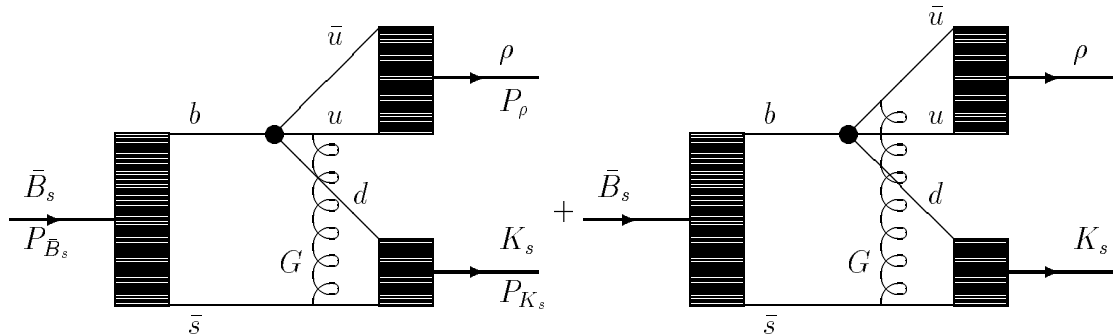}
\end{center}
\vspace{ -1.00cm}
\caption{\baselineskip=7mm     The colour exchange
graphs for the process $\bar B_s
           \rightarrow           \rho
                 +       K_S$ to leading order in the
Lepage-Brodsky expansion in Ref.~\cite{L-B,shbrod}, ignoring penguins.
The kinematics is as defined in Fig.~\ref{figone}.
}
\label{figtwo}
\end{figure}
%------------begin fig 3 --------------------------
\begin{figure}
\begin{center}
\epsfig{file=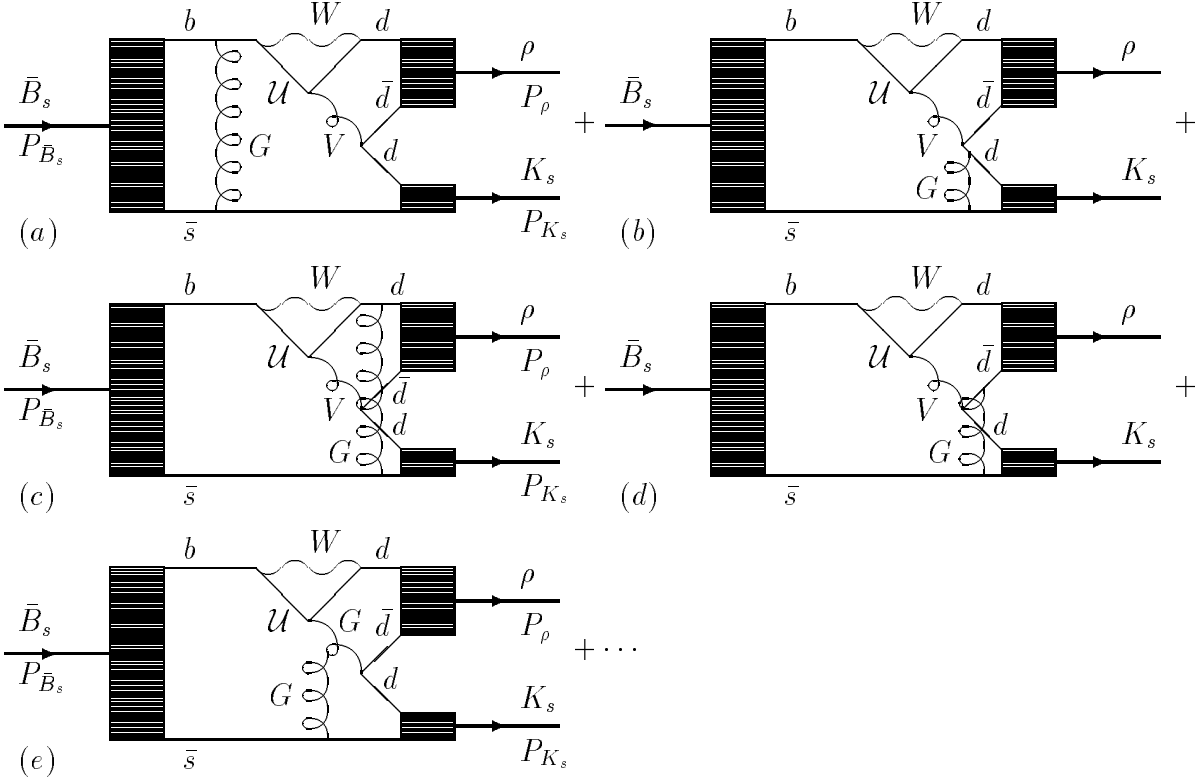}
\end{center}
\vspace{ -1.00cm}
\caption{\baselineskip=7mm     The penguin graphs for the process $\bar B_s
           \rightarrow           \rho
                 +       K_S$, to leading order in the
Lepage-Brodsky expansion defined in Ref.~\cite{L-B,shbrod}.
The kinematics is as defined in Fig.~\ref{figone}.
}
\label{figthree}
\end{figure}
\noindent
In Fig.~\ref{figtwo}, we show the graphs in which colour is exchanged
between the would-be spectator $\bar s$ in Fig.~\ref{figone} and the
outgoing $\rho$ parton lines and in Fig.~\ref{figthree} we show
the respective penguin graphs: the dominant graphs according to
the prescription in Ref.~\cite{shbrod}
(3a,3b), the colour exchange graphs (3c,3d), and the exchange of the hard
gluon $G$ between the would-be spectator $\bar s$ and the penguin gluon 
itself for QCD penguins, (3e), which we also will classify as colour exchange.
To address the issue of factorisation/colour-suppression,
we shall present results when graphs in Figs. 2 and 3c-3e
are dropped and when they are included.
We thus give results for
the approximations in which only the graphs in Fig.~\ref{figone}
are included (Tree), in which the graphs in Figs. 1, 
3a and 3b are included (Tree$+$Penguin), in which
the graphs in Figs. 1,2,3a and 3b are included
(Tree$+$Penguin$+$Tree Colour Exchange($CE_T$)), and in which
all graphs in Figs. 1,2,3a-3e are included 
(Tree$+$Penguin$+$Tree and 
Penguin Colour Exchange($CE_{T+P}$)).
For the electroweak (EW) penguins, there is no penguin gluon with which
the would-be spectator $\bar s$ could interact.
We need to stress that, as shown in Ref.~\cite{shbrod},
the usual QCD factorisation properties for exclusive
amplitudes at large momentum transfer
are sufficient to justify the formulation of our amplitude
according to the graphs in Figs. 1-3. More phenomenological
arguments, such as the current field identity based BSW model
in Ref.~\cite{BSW}, etc.,
which would lead to the same graphs, are not needed.\par
Some discussion of the effective values of the 
coefficients $C_1=\tilde C_2,~C_2=\tilde C_1$ in relation to 
the coefficients $a_1$ and $a_2$
as defined in Ref.~\cite{BSW} is now appropriate. 
Following Ref.~\cite{BSW}
and the recent results in Ref.~\cite{BROWDER}, 
when we use the standard QCD to calculate the
diagrams in Fig. 1 and take them alone as our
result ( this is our definition of factorisation)
, we use $a_2\cong .24\cong |C_2(m_b)|$
and when we assess the colour-suppression effect by including
the exchange of $G$ between the $\bar s$ and the $q\bar q$ of the
$\rho$ we set $C_1(m_b)\cong 1.1$; 
these results are consistent with those found in Ref.~\cite{BROWDER}.
We note that the naive relation
$a_2\cong C_2+{1\over 3}C_1\cong .127$ would give a value for $a_2$
that is about a factor of two smaller than what is found in
Refs.~\cite{BSW,BROWDER} and the references therein. The parameters
$a_1,a_2$ are therefore purely phenomenological properties of
the hard effective weak interaction process and can be taken from
experiment in our analysis: one may view $a_2$, for example, as the
effective value of $C_2+{1\over3}C_1$ when the current field identity
is used to interpolate the $\rho$ into our effective weak interaction
vertex. The Lepage-Brodsky formalism then allows us to calculate
the recoil corrections associated with the momentum transfer required
for the would-be spectator to be kicked from the $B_s$ to the
final outgoing $K_S$ using perturbative QCD to describe the respective
hard gluon exchange, as we noted above. This ``kick'' is the defining
aspect of our calculation of $\bar B_s\rightarrow \rho K_S$
in comparison to those in Refs.~\cite{others1} and in fact in
comparison to the related two body B decay analyses in Refs.~\cite{others2}.
The point is the following. As one can see from the results in
Refs.~\cite{pipi,dppkk,ccarlsn}, contrary to what happens in the
tree level part of the calculations in Refs.~\cite{others1,others2},
the graph in Fig.~1a in which the hard gluon kick to the spectator
comes from the b-quark line (the heavy quark line) develops
an imaginary part that is treated rigorously in our work
so that there is a non-trivial strong phase for our ``tree level''
contribution compared to those in Refs.~\cite{others1,others2}. 
This happens because, as $m_B > m_b+m_s$ where 
$m_q$ are evaluated at the scale $\sim m_B$, the heavy quark line
can reach its perturbative QCD mass shell in the graph in Fig.~1a,
and in the similar graphs in Figs.~2 and 3. Evidently, this effect
is missing in the results in Refs.~\cite{others1,others2}. Any serious
discussion of the CP asymmetries in the amplitudes for exclusive
two-body B decays must take this strong phase into account in general
(it is different for Tree and Penguin contributions for example)
as one can see from our formula for time-dependent asymmetry
in $ \bar B_s\rightarrow \rho K_S$ below. Our paper is the 
first paper to do this systematically.\par
Here, we should also comment on the recent results of Ref.~\cite{beneke}
on the process $B\rightarrow \pi\pi$. The authors in Ref.~\cite{beneke}
use the same Feynman diagrams, analogous to those
in Figs. 1-3 here, as we have shown already in
Ref.~\cite{pipi} and  same Lepage-Brodsky expansion formalism
except that they assume the graphs analogous to those in Fig. 1
are to be replaced by a real form factor with the appropriate
external wave function/decay constant factors.
The usual corrections
to the diagrams in Fig. 1 are then represented as a power series
in $\alpha_s$ times this assumed real form factor. We do not make such an
assumption; we calculate systematically in the Lepage-Brodsky
expansion. A major difference is that the authors in Ref.~\cite{beneke}
miss the recoil phase of the dominant contribution in the 
analog of Fig.~1 for the $B\rightarrow \pi\pi$ process,
although they do calculate the recoil phase in the respective analogs
of Figs. 2 and 3. To see what effect this has,
we note that, from our Eq.(5) in Ref.~\cite{pipi}, we get the
direct CP violation result~\cite{th-2000-101}
for the $\bar B\rightarrow \pi\pi$ process as
\begin{equation}
-0.0086 < {\cal A}^{dir}_{CP} < 0,~\text{for} ~\gamma\in(0,\pi)\cr
-0.0086 < {\cal A}^{dir}_{CP} < -0.0050,~\text{for}~\frac{3\pi}{4}\ge\gamma
\ge\frac{\pi}{4},
\label{adir}
\end{equation}
with ${\cal A}^{dir}_{CP}$ monotone decreasing in the second
currently preferred~\cite{sbfacwkp} region of $\gamma$
for $\frac{\pi}{4}\le \gamma\le 1.806$ and monotone increasing for 
$1.806\le \gamma \le \frac{3\pi}{4}$,
whereas in Ref.~\cite{beneke} this asymmetry is predicted to be
$-4\%\times\sin\gamma$. Evidently, experiment will soon be able to
distinguish between the two approaches. See Ref.~\cite{elsewhere}
for further discussion of this and related matters.
\par  
In this way, using the methods of Ref.~\cite{L-B} we evaluate
the graphs illustrated in Fig.~1-3 and arrive at the results in Table 1
and in Fig. 4 ( the explicit expressions for the
respective amplitudes may be inferred from those
for the process $\bar B\rightarrow \pi\pi$ given in Eq.(1)
and in Eqs.(A1-A4) in
Ref.~\cite{pipi} via the appropriate substitutions of 
momenta and distribution amplitudes; for example, for the factor
$F_N$ in (A1)in Ref.~\cite{pipi} we would now have its form 
obtained by the substitutions
\begin{align}
a_1 &\rightarrow a_2 \notag\\
\sqrt{2}f_\pi P_{{\pi^-}\alpha}&\rightarrow f_\rho m_\rho\epsilon(P_\rho)_\alpha\notag\\
{\sqrt{3}f_\pi\over 2}y_1y_2&\rightarrow {\sqrt{3}f_K\over 2\sqrt{2}}y_1y_2
(1+3{\beta'}_K(y_2-y_1))\notag\\
P_{\pi^+}&\rightarrow P_{K_S}\notag\\
m_u&\rightarrow m_d\notag\\
m_\pi&\rightarrow m_{K_S}\notag\\
4x_2^2y_2E_{\pi^+}^2m_B^2&\rightarrow q^2[(P_d+q)^2-m_d^2+i\epsilon]\notag\\
(-2x_2y_2E_{\pi^+}m_B)(y_1m_B^2-m_b^2+i\epsilon)&\rightarrow q^2[(P_b-q)^2-m_b^2+i\epsilon]\notag\\
\end{align} 
wherein $q=P_{\bar s}'-P_{\bar s}$, $P_f$,~ $f=b,~\bar s$ is
4-momentum of $f$ in the $\bar B_s$ in Fig.~1 and $P_d,P_{\bar s}'$
are the 4-momentum of $d,\bar s$ respectively in the $K_S$ in Fig.~1, so that
we have $P_{\bar s}\cong x_2P_{\bar B_s}$ and $P_{\bar s}'\cong y_2P_{K_S}$,
for example, and ${\beta'}_K\cong .418$ is the asymmetry parameter
in the Lepage-Brodsky distribution amplitude for the $K_S$ 
as determined in Ref.~\cite{semdcy} and evolved to the scale $m_B$.
We use $f_K\cong 0.112$.
In this regard, we further note that the Lepage-Brodsky distribution
amplitude for the $\rho$ in the analog of Eq.(A4) in Ref.~\cite{pipi} for
the process under study here would substitute $\sqrt{3}f_\rho m_\rho
\not\!\epsilon(P_\rho)z_1z_2(1.348-1.74z_1+1.74z_1^2)$ for \\
$\sqrt{3}f_\pi z_1z_2\gamma_5(\not\!P'_{\pi^o}+m_\pi)$ for example
by the standard methods, where we use the Chernyak-Zhitnitsky (C-Z)
type result~\cite{C-Z} for
the $\rho$ distribution amplitude in analogy with our discussion
in the Notes Added in Ref.~\cite{pipi}. Here, $\epsilon(P_\rho)$
and $f_\rho$ are the respective $\rho$ polarisation
4-vector and decay constant with $f_\rho\cong .14GeV$. The $\bar B_s$
distribution amplitude
is taken in complete analogy with the $\bar B_d$ in Ref.~\cite{pipi},
so that it is given by 
$a_B\phi_B(w_1,w_2)/\sqrt{2N_c}=a_B\delta(w_2-x_2)/\sqrt{2N_c}$
where $N_c=3$ is the number of colours, $a_B=f_{B_s}/\sqrt{4N_c}$
and $x_2\cong 0.0542$ is determined, as we present in our
Appendix, following the treatment of heavy mesons in
Ref.~\cite{L-B} using potential model parameters such as
those in Ref.~\cite{eichten:1980}.
Finally, note that the quark masses $m_q$ are the running current
quark masses~\cite{masses}).
For completeness, the complete result for the amplitude 
corresponding to the graphs shown in Figs.~1-3
is given in the Appendix. Moreover, the precise definition of
the penguin shift $\Delta \sin(2\gamma)$ is given by the following
generalisation to our process $\bar B_s\rightarrow \rho K_S$
of the formula of Gronau in Ref.~\cite{gronau} for the corresponding
shift of $\sin(2\alpha)$ due to penguins in the $\bar B_d\rightarrow \pi\pi$
process
\begin{equation}
-\sin(2\gamma)-\Delta(\sin(2\gamma))\equiv {\Im \lambda \over {1\over2}(1+|\lambda|^2)} 
\label{pshift}
\end{equation}
for
\begin{equation}
\lambda={A_Te^{-i\phi_T+i\delta_T}+\sum_j A_{P_j}e^{-i\phi_{P_j}+i\delta_{P_j}}
\over A_Te^{+i\phi_T+i\delta_T}+\sum_j
A_{P_j}e^{+i\phi_{P_j}+i\delta_{P_j}}},
\label{lambda}
\end{equation} 
where the amplitude $A_Te^{-i\phi_T+i\delta_T}$ corresponds to the 
tree-level weak processes
in Figs. 1 and 2  and the amplitudes 
$ A_{P_j}e^{-i\phi_{P_j}+i\delta_{P_j}}$ correspond to the respective penguin
processes in Fig. 3. Here, we identify the weak phases of the respective
amplitudes as $\phi_r$,~$r=T,P_j$ and the attendant strong phases as
$\delta_r$,~$r=T,P_j$. In general, $j=1,2$ distinguishes the electric
and magnetic penguins when this is required, as one can see in our
Appendix. In this notation, we have $\gamma\equiv \phi_T$.
%%%%begin table 1%%%%%%%%%%%%%%%%%%%%%%%%%%%%%%%%%%%%%%%%%%%%%%%%%%%%%%%%%%%%
%%-----------
\begin{table}
\centering
\newcommand{\lstrut}{{$\strut\atop\strut$}}
\newcommand{\alpi}{\big({\alpha\over\pi}\big)}
\newcommand{\B}{\Big}

\vspace{0.5cm}
\begin{tabular}{|l|l|l|l|l|}
\hline
\multicolumn{5}{|c|}{ $BR(\bar B\rightarrow \rho K_s)/((f_{B_s}/.141GeV)^2) $ }
\\  \hline
\cline{2-5}
&$Tree$& $Tree+Penguin$ & $Tree+Penguin+CE_T$&$Tree+Penguin+CE_{T+P}$\\ \hline
%{$R_b$} &\multicolumn{3}{|c|}\hfil \\ \hline
{$R_b$} &\multicolumn{4}{|c|}{$10^{-8}$} \\ \hline
\cline{2-5}
%%$0.220$ &$0.0352$&$[0.0296,0.0875]$&$[0.00326,0.169]$&$[0.00590,0.547]$
$0.220$ &$0.0352$&$[0.0296,0.0875]$&$[0.0111,0.823]$&$[0.000205,0.646]$
%%                             &\multicolumn{2}{|c|}{$^{+.15\%}_{-.065\%}$}
\\ \hline
\cline{2-5}
%%$0.385$&$0.108$&$[0.0158,0.117]$&$[0.00574,0.295]$&$[0.0297,0.978]$
$0.385$&$0.108$&$[0.0158,0.117]$&$[0.236,1.66]$&$[0.0805,1.21]$
\\ \hline
%Born
%{Born}&\multicolumn{3}{|c|}\hfil  \\ \hline
\cline{2-5}
%%$0.551$  & $0.221$&$[0.00624,0.151]$&$[0.0436,0.458]$&$[0.178,1.53]$
$0.551$  & $0.221$&$[0.00624,0.151]$&$[0.752,2.79]$&$[0.338,1.95]$
%\\ \cline{2-4}
%$2!\sigma_{trig}(nb)$&$5140.57\pm1.08$&$3395.41\pm.68$&$2509.74\pm.75$\\
%%%                        \multicolumn{2}{|c|}{ $<0.002\%$  }
%\\ \cline{2-3}
%\{7\} pairs NLL, quarks, $\mu$ &
%                        \multicolumn{2}{|c|}{ $<0.007\%$  }
%\\ \cline{2-3}
%\{8\} $Z$ exchange        &\multicolumn{2}{|c|}{ $0.033\%$ }\\
%\cline{2-3}
%\{9\} $\gamma$ exchange $s$-channel&\multicolumn{2}{|c|}{$<0.0002\%$ }\\
%\hline
%Total                    &\multicolumn{2}{|c|}{$^{+.16\%}_{-.089\%}$}\\
\\
\hline
\end{tabular}
\caption{\it  BR for $\protect\bar B_s\rightarrow \rho K_S$
as a function of $\protect R_b$ as defined in the text. The factorised
approximation without penguin effects is denoted as $Tree$;
the corresponding results with the penguin effects (both EW and QCD penguins)
included are denoted by $Tree+Penguin$; the results corresponding
to the inclusion of the gluon exchange between the $\protect u\bar u$
in the $\protect\rho$ and the $\protect\bar s$ would-be spectator
are denoted by $Tree+Penguin+CE_T$; and, when the gluon exchanges
between the $\protect\bar s$ would-be spectator and the outgoing
$\protect d\bar d$ of the $\rho$ and the penguin gluon itself are included,
we denote the result by $Tree+Penguin+CE_{T+P}$.
All results are given 
with a factor of $(f_{B_s}/.141GeV)^2\times 10^{-8}$ removed for a total width
$\protect\Gamma(B_s\rightarrow all)=4.085\times 10^{-13}GeV$ and for
the variation $0.0\le\delta_{13}\le 2\pi$.
}
\end{table}
%%----------------
%%%%%%%%%%%%%%%%%%end table 1%%%%%%%%%%%%%%%%%%%%%%%%%%%%%%%%%%%%%%%%%%
%%%%%%%%%%begin fig. 4%%%%%%%%%%%%%%%%%%%%%%%%%%%%%%%%%%%%%%%%%%%%%%%%
%\voffset =  1.0cm
%\hoffset = -1cm
%\documentstyle[12pt]{article}
%\textwidth  = 16cm
%\textheight = 24cm
%\begin{document}

% =========== big frame, title etc. =======
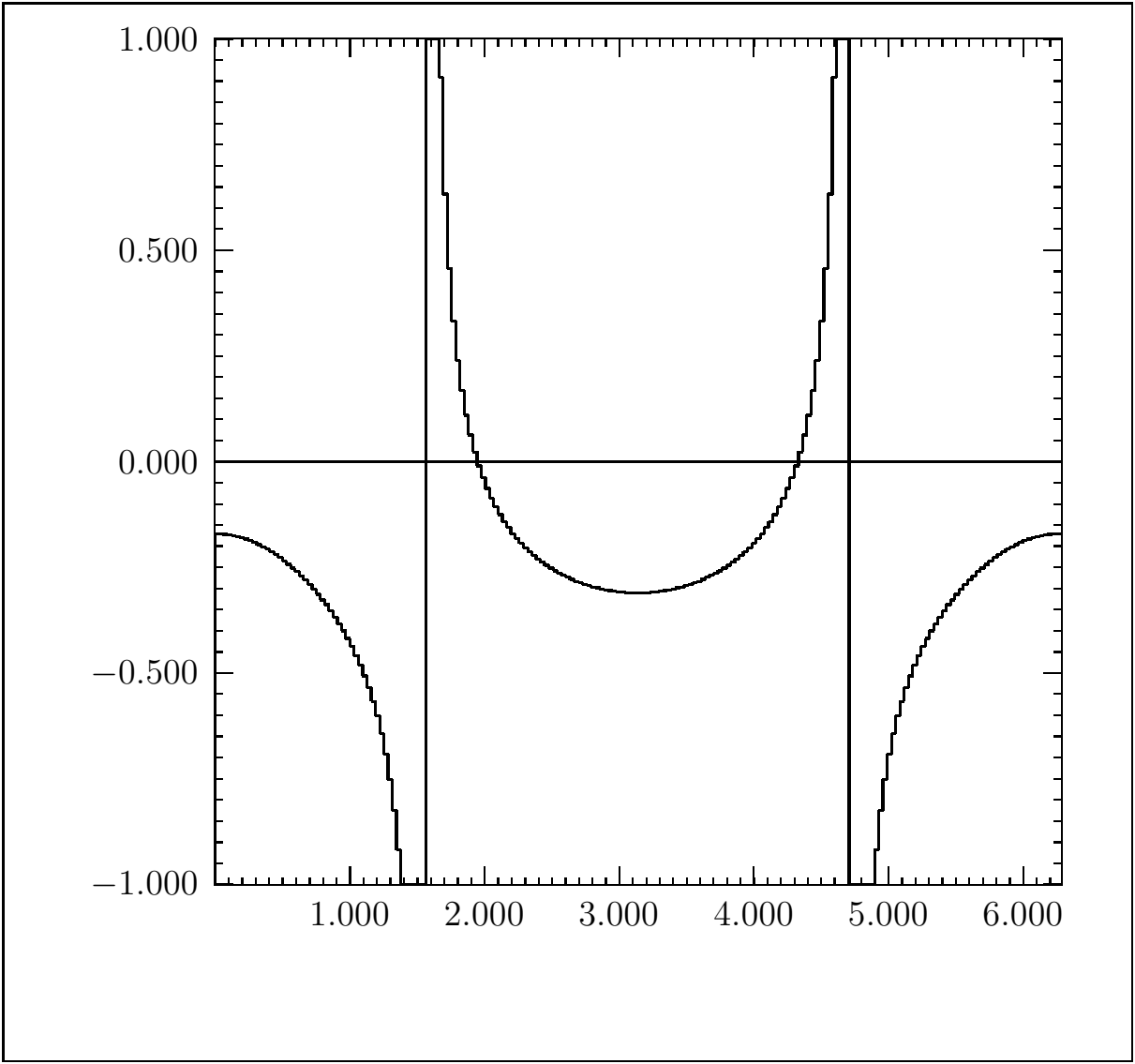
\begin{figure}
\begin{center}
  Penguin Shift of $\sin(2\gamma)$                                             
\end{center}
\setlength{\unitlength}{0.1mm}
\begin{picture}(1600,1500)
\put(0,0){\framebox(1600,1500){ }}
% =========== small frame, labeled axis ===
\put(300,250){\begin{picture}( 1200,1200)
\put(0,0){\framebox( 1200,1200){ }}
% =========== x and y axis ================
% .......SAXIX........ 
%  JY=    6
\multiput(  190.99,0)(  190.99,0){   6}{\line(0,1){25}}
\multiput(     .00,0)(   19.10,0){  63}{\line(0,1){10}}
\multiput(  190.99,1200)(  190.99,0){   6}{\line(0,-1){25}}
\multiput(     .00,1200)(   19.10,0){  63}{\line(0,-1){10}}
\put( 190,-25){\makebox(0,0)[t]{\large $    1.000 $}}
\put( 381,-25){\makebox(0,0)[t]{\large $    2.000 $}}
\put( 572,-25){\makebox(0,0)[t]{\large $    3.000 $}}
\put( 763,-25){\makebox(0,0)[t]{\large $    4.000 $}}
\put( 954,-25){\makebox(0,0)[t]{\large $    5.000 $}}
\put(1145,-25){\makebox(0,0)[t]{\large $    6.000 $}}
% .......SAXIY........ 
%  JY=    2
\multiput(0,     .00)(0,  300.00){   5}{\line(1,0){25}}
\multiput(0,     .00)(0,   30.00){  41}{\line(1,0){10}}
\multiput(1200,     .00)(0,  300.00){   5}{\line(-1,0){25}}
\multiput(1200,     .00)(0,   30.00){  41}{\line(-1,0){10}}
\put(0,  600.00){\line(1,0){1200}}
\put(-25,   0){\makebox(0,0)[r]{\large $  -1.000 $}}
\put(-25, 300){\makebox(0,0)[r]{\large $   -0.500 $}}
\put(-25, 600){\makebox(0,0)[r]{\large $    0.000 $}}
\put(-25, 900){\makebox(0,0)[r]{\large $    0.500 $}}
\put(-25,1200){\makebox(0,0)[r]{\large $   1.000 $}}
\end{picture}}% end of plotting labeled axis
%========== next plot (line) ==========
%==== HISTOGRAM ID=   900
%  Delta-Gamma in PerCent                                               
\put(300,250){\begin{picture}( 1200,1200)
% ========== plotting primitives ==========
\thicklines 
\newcommand{\x}[3]{\put(#1,#2){\line(1,0){#3}}}
\newcommand{\y}[3]{\put(#1,#2){\line(0,1){#3}}}
\newcommand{\z}[3]{\put(#1,#2){\line(0,-1){#3}}}
\newcommand{\e}[3]{\put(#1,#2){\line(0,1){#3}}}
\y{   0}{   0}{ 498}\x{   0}{ 498}{   5}
\y{   5}{ 498}{   0}\x{   5}{ 498}{   6}
\z{  11}{ 498}{   1}\x{  11}{ 497}{   6}
\z{  17}{ 497}{   1}\x{  17}{ 496}{   6}
\z{  23}{ 496}{   1}\x{  23}{ 495}{   6}
\z{  29}{ 495}{   1}\x{  29}{ 494}{   6}
\z{  35}{ 494}{   2}\x{  35}{ 492}{   6}
\z{  41}{ 492}{   2}\x{  41}{ 490}{   6}
\z{  47}{ 490}{   2}\x{  47}{ 488}{   6}
\z{  53}{ 488}{   3}\x{  53}{ 485}{   6}
\z{  59}{ 485}{   3}\x{  59}{ 482}{   6}
\z{  65}{ 482}{   3}\x{  65}{ 479}{   6}
\z{  71}{ 479}{   3}\x{  71}{ 476}{   6}
\z{  77}{ 476}{   4}\x{  77}{ 472}{   6}
\z{  83}{ 472}{   4}\x{  83}{ 468}{   6}
\z{  89}{ 468}{   4}\x{  89}{ 464}{   6}
\z{  95}{ 464}{   5}\x{  95}{ 459}{   6}
\z{ 101}{ 459}{   5}\x{ 101}{ 454}{   6}
\z{ 107}{ 454}{   5}\x{ 107}{ 449}{   6}
\z{ 113}{ 449}{   5}\x{ 113}{ 444}{   6}
\z{ 119}{ 444}{   6}\x{ 119}{ 438}{   6}
\z{ 125}{ 438}{   6}\x{ 125}{ 432}{   6}
\z{ 131}{ 432}{   6}\x{ 131}{ 426}{   6}
\z{ 137}{ 426}{   7}\x{ 137}{ 419}{   6}
\z{ 143}{ 419}{   7}\x{ 143}{ 412}{   6}
\z{ 149}{ 412}{   8}\x{ 149}{ 404}{   6}
\z{ 155}{ 404}{   7}\x{ 155}{ 397}{   6}
\z{ 161}{ 397}{   9}\x{ 161}{ 388}{   6}
\z{ 167}{ 388}{   9}\x{ 167}{ 379}{   6}
\z{ 173}{ 379}{   9}\x{ 173}{ 370}{   6}
\z{ 179}{ 370}{  10}\x{ 179}{ 360}{   6}
\z{ 185}{ 360}{  11}\x{ 185}{ 349}{   6}
\z{ 191}{ 349}{  11}\x{ 191}{ 338}{   6}
\z{ 197}{ 338}{  13}\x{ 197}{ 325}{   6}
\z{ 203}{ 325}{  14}\x{ 203}{ 311}{   6}
\z{ 209}{ 311}{  15}\x{ 209}{ 296}{   6}
\z{ 215}{ 296}{  17}\x{ 215}{ 279}{   6}
\z{ 221}{ 279}{  19}\x{ 221}{ 260}{   6}
\z{ 227}{ 260}{  21}\x{ 227}{ 239}{   6}
\z{ 233}{ 239}{  25}\x{ 233}{ 214}{   6}
\z{ 239}{ 214}{  29}\x{ 239}{ 185}{   6}
\z{ 245}{ 185}{  36}\x{ 245}{ 149}{   6}
\z{ 251}{ 149}{  44}\x{ 251}{ 105}{   6}
\z{ 257}{ 105}{  56}\x{ 257}{  49}{   6}
\z{ 263}{  49}{  49}\x{ 263}{   0}{   6}
\y{ 269}{   0}{   0}\x{ 269}{   0}{   6}
\y{ 275}{   0}{   0}\x{ 275}{   0}{   6}
\y{ 281}{   0}{   0}\x{ 281}{   0}{   6}
\y{ 287}{   0}{   0}\x{ 287}{   0}{   6}
\y{ 293}{   0}{   0}\x{ 293}{   0}{   6}
\y{ 299}{   0}{1200}\x{ 299}{1200}{   6}
\y{ 305}{1200}{   0}\x{ 305}{1200}{   6}
\y{ 311}{1200}{   0}\x{ 311}{1200}{   6}
\z{ 317}{1200}{  54}\x{ 317}{1146}{   6}
\z{ 323}{1146}{ 166}\x{ 323}{ 980}{   6}
\z{ 329}{ 980}{ 106}\x{ 329}{ 874}{   6}
\z{ 335}{ 874}{  75}\x{ 335}{ 799}{   6}
\z{ 341}{ 799}{  55}\x{ 341}{ 744}{   6}
\z{ 347}{ 744}{  43}\x{ 347}{ 701}{   6}
\z{ 353}{ 701}{  35}\x{ 353}{ 666}{   6}
\z{ 359}{ 666}{  28}\x{ 359}{ 638}{   6}
\z{ 365}{ 638}{  24}\x{ 365}{ 614}{   6}
\z{ 371}{ 614}{  20}\x{ 371}{ 594}{   6}
\z{ 377}{ 594}{  17}\x{ 377}{ 577}{   6}
\z{ 383}{ 577}{  15}\x{ 383}{ 562}{   6}
\z{ 389}{ 562}{  14}\x{ 389}{ 548}{   6}
\z{ 395}{ 548}{  12}\x{ 395}{ 536}{   6}
\z{ 401}{ 536}{  11}\x{ 401}{ 525}{   6}
\z{ 407}{ 525}{  10}\x{ 407}{ 515}{   6}
\z{ 413}{ 515}{   8}\x{ 413}{ 507}{   6}
\z{ 419}{ 507}{   9}\x{ 419}{ 498}{   6}
\z{ 425}{ 498}{   7}\x{ 425}{ 491}{   6}
\z{ 431}{ 491}{   7}\x{ 431}{ 484}{   6}
\z{ 437}{ 484}{   6}\x{ 437}{ 478}{   6}
\z{ 443}{ 478}{   6}\x{ 443}{ 472}{   6}
\z{ 449}{ 472}{   5}\x{ 449}{ 467}{   6}
\z{ 455}{ 467}{   5}\x{ 455}{ 462}{   6}
\z{ 461}{ 462}{   5}\x{ 461}{ 457}{   6}
\z{ 467}{ 457}{   4}\x{ 467}{ 453}{   6}
\z{ 473}{ 453}{   4}\x{ 473}{ 449}{   6}
\z{ 479}{ 449}{   4}\x{ 479}{ 445}{   6}
\z{ 485}{ 445}{   3}\x{ 485}{ 442}{   6}
\z{ 491}{ 442}{   3}\x{ 491}{ 439}{   6}
\z{ 497}{ 439}{   3}\x{ 497}{ 436}{   6}
\z{ 503}{ 436}{   3}\x{ 503}{ 433}{   6}
\z{ 509}{ 433}{   3}\x{ 509}{ 430}{   6}
\z{ 515}{ 430}{   2}\x{ 515}{ 428}{   6}
\z{ 521}{ 428}{   2}\x{ 521}{ 426}{   6}
\z{ 527}{ 426}{   2}\x{ 527}{ 424}{   6}
\z{ 533}{ 424}{   2}\x{ 533}{ 422}{   6}
\z{ 539}{ 422}{   1}\x{ 539}{ 421}{   6}
\z{ 545}{ 421}{   2}\x{ 545}{ 419}{   6}
\z{ 551}{ 419}{   1}\x{ 551}{ 418}{   6}
\z{ 557}{ 418}{   1}\x{ 557}{ 417}{   6}
\z{ 563}{ 417}{   1}\x{ 563}{ 416}{   6}
\z{ 569}{ 416}{   1}\x{ 569}{ 415}{   6}
\y{ 575}{ 415}{   0}\x{ 575}{ 415}{   6}
\z{ 581}{ 415}{   1}\x{ 581}{ 414}{   6}
\y{ 587}{ 414}{   0}\x{ 587}{ 414}{   6}
\y{ 593}{ 414}{   0}\x{ 593}{ 414}{   6}
\y{ 599}{ 414}{   0}\x{ 599}{ 414}{   6}
\y{ 605}{ 414}{   0}\x{ 605}{ 414}{   6}
\y{ 611}{ 414}{   0}\x{ 611}{ 414}{   6}
\y{ 617}{ 414}{   1}\x{ 617}{ 415}{   6}
\y{ 623}{ 415}{   0}\x{ 623}{ 415}{   6}
\y{ 629}{ 415}{   1}\x{ 629}{ 416}{   6}
\y{ 635}{ 416}{   1}\x{ 635}{ 417}{   6}
\y{ 641}{ 417}{   1}\x{ 641}{ 418}{   6}
\y{ 647}{ 418}{   1}\x{ 647}{ 419}{   6}
\y{ 653}{ 419}{   2}\x{ 653}{ 421}{   6}
\y{ 659}{ 421}{   1}\x{ 659}{ 422}{   6}
\y{ 665}{ 422}{   2}\x{ 665}{ 424}{   6}
\y{ 671}{ 424}{   2}\x{ 671}{ 426}{   6}
\y{ 677}{ 426}{   2}\x{ 677}{ 428}{   6}
\y{ 683}{ 428}{   2}\x{ 683}{ 430}{   6}
\y{ 689}{ 430}{   3}\x{ 689}{ 433}{   6}
\y{ 695}{ 433}{   3}\x{ 695}{ 436}{   6}
\y{ 701}{ 436}{   3}\x{ 701}{ 439}{   6}
\y{ 707}{ 439}{   3}\x{ 707}{ 442}{   6}
\y{ 713}{ 442}{   3}\x{ 713}{ 445}{   6}
\y{ 719}{ 445}{   4}\x{ 719}{ 449}{   6}
\y{ 725}{ 449}{   4}\x{ 725}{ 453}{   6}
\y{ 731}{ 453}{   4}\x{ 731}{ 457}{   6}
\y{ 737}{ 457}{   5}\x{ 737}{ 462}{   6}
\y{ 743}{ 462}{   5}\x{ 743}{ 467}{   6}
\y{ 749}{ 467}{   5}\x{ 749}{ 472}{   6}
\y{ 755}{ 472}{   6}\x{ 755}{ 478}{   6}
\y{ 761}{ 478}{   6}\x{ 761}{ 484}{   6}
\y{ 767}{ 484}{   7}\x{ 767}{ 491}{   6}
\y{ 773}{ 491}{   7}\x{ 773}{ 498}{   6}
\y{ 779}{ 498}{   9}\x{ 779}{ 507}{   6}
\y{ 785}{ 507}{   8}\x{ 785}{ 515}{   6}
\y{ 791}{ 515}{  10}\x{ 791}{ 525}{   6}
\y{ 797}{ 525}{  11}\x{ 797}{ 536}{   6}
\y{ 803}{ 536}{  12}\x{ 803}{ 548}{   6}
\y{ 809}{ 548}{  14}\x{ 809}{ 562}{   6}
\y{ 815}{ 562}{  15}\x{ 815}{ 577}{   6}
\y{ 821}{ 577}{  17}\x{ 821}{ 594}{   6}
\y{ 827}{ 594}{  20}\x{ 827}{ 614}{   6}
\y{ 833}{ 614}{  24}\x{ 833}{ 638}{   6}
\y{ 839}{ 638}{  28}\x{ 839}{ 666}{   6}
\y{ 845}{ 666}{  35}\x{ 845}{ 701}{   6}
\y{ 851}{ 701}{  43}\x{ 851}{ 744}{   6}
\y{ 857}{ 744}{  55}\x{ 857}{ 799}{   6}
\y{ 863}{ 799}{  75}\x{ 863}{ 874}{   6}
\y{ 869}{ 874}{ 106}\x{ 869}{ 980}{   6}
\y{ 875}{ 980}{ 166}\x{ 875}{1146}{   6}
\y{ 881}{1146}{  54}\x{ 881}{1200}{   6}
\y{ 887}{1200}{   0}\x{ 887}{1200}{   6}
\y{ 893}{1200}{   0}\x{ 893}{1200}{   6}
\z{ 899}{1200}{1200}\x{ 899}{   0}{   6}
\y{ 905}{   0}{   0}\x{ 905}{   0}{   6}
\y{ 911}{   0}{   0}\x{ 911}{   0}{   6}
\y{ 917}{   0}{   0}\x{ 917}{   0}{   6}
\y{ 923}{   0}{   0}\x{ 923}{   0}{   6}
\y{ 929}{   0}{   0}\x{ 929}{   0}{   6}
\y{ 935}{   0}{  49}\x{ 935}{  49}{   6}
\y{ 941}{  49}{  56}\x{ 941}{ 105}{   6}
\y{ 947}{ 105}{  44}\x{ 947}{ 149}{   6}
\y{ 953}{ 149}{  36}\x{ 953}{ 185}{   6}
\y{ 959}{ 185}{  29}\x{ 959}{ 214}{   6}
\y{ 965}{ 214}{  25}\x{ 965}{ 239}{   6}
\y{ 971}{ 239}{  21}\x{ 971}{ 260}{   6}
\y{ 977}{ 260}{  19}\x{ 977}{ 279}{   6}
\y{ 983}{ 279}{  17}\x{ 983}{ 296}{   6}
\y{ 989}{ 296}{  15}\x{ 989}{ 311}{   6}
\y{ 995}{ 311}{  14}\x{ 995}{ 325}{   6}
\y{1001}{ 325}{  13}\x{1001}{ 338}{   6}
\y{1007}{ 338}{  11}\x{1007}{ 349}{   6}
\y{1013}{ 349}{  11}\x{1013}{ 360}{   6}
\y{1019}{ 360}{  10}\x{1019}{ 370}{   6}
\y{1025}{ 370}{   9}\x{1025}{ 379}{   6}
\y{1031}{ 379}{   9}\x{1031}{ 388}{   6}
\y{1037}{ 388}{   9}\x{1037}{ 397}{   6}
\y{1043}{ 397}{   7}\x{1043}{ 404}{   6}
\y{1049}{ 404}{   8}\x{1049}{ 412}{   6}
\y{1055}{ 412}{   7}\x{1055}{ 419}{   6}
\y{1061}{ 419}{   7}\x{1061}{ 426}{   6}
\y{1067}{ 426}{   6}\x{1067}{ 432}{   6}
\y{1073}{ 432}{   6}\x{1073}{ 438}{   6}
\y{1079}{ 438}{   6}\x{1079}{ 444}{   6}
\y{1085}{ 444}{   5}\x{1085}{ 449}{   6}
\y{1091}{ 449}{   5}\x{1091}{ 454}{   6}
\y{1097}{ 454}{   5}\x{1097}{ 459}{   6}
\y{1103}{ 459}{   5}\x{1103}{ 464}{   6}
\y{1109}{ 464}{   4}\x{1109}{ 468}{   6}
\y{1115}{ 468}{   4}\x{1115}{ 472}{   6}
\y{1121}{ 472}{   4}\x{1121}{ 476}{   6}
\y{1127}{ 476}{   3}\x{1127}{ 479}{   6}
\y{1133}{ 479}{   3}\x{1133}{ 482}{   6}
\y{1139}{ 482}{   3}\x{1139}{ 485}{   6}
\y{1145}{ 485}{   3}\x{1145}{ 488}{   6}
\y{1151}{ 488}{   2}\x{1151}{ 490}{   6}
\y{1157}{ 490}{   2}\x{1157}{ 492}{   6}
\y{1163}{ 492}{   2}\x{1163}{ 494}{   6}
\y{1169}{ 494}{   1}\x{1169}{ 495}{   6}
\y{1175}{ 495}{   1}\x{1175}{ 496}{   6}
\y{1181}{ 496}{   1}\x{1181}{ 497}{   6}
\y{1187}{ 497}{   1}\x{1187}{ 498}{   6}
\y{1193}{ 498}{   0}\x{1193}{ 498}{   6}
\end{picture}} % end of plotting histogram
\end{picture} % close entire picture 
\label{figfour}
\caption{ \baselineskip=7mm
Penguin shift of the CP asymmetry $\sin(2\gamma)$ in
$\bar B_s\rightarrow \rho K_S$ for $R_b=0.385$ for the matrix element
approximation corresponding to the last column in Table 1. The analogous plots
for the $\pm1\sigma$ values of $R_b$ are discussed in the text.}
\end{figure}
%%%%%%%%%%end fig.4%%%%%%%%%%%%%%%%%%%%%%%%%%%%%%%%%%%%%%%%%%%%%%%%%%%
\noindent
From the results in Table 1, and their ratios with one another,
we see that the colour suppression idea does
not really hold for this decay. We see, as 
already anticipated by several authors~\cite{sbfacwkp},
that the penguins are indeed important. 
There is
a regime,~
% $0.0^o\leq \gamma \lesssim 24.8^o$,~ 
$0.0^o\leq \gamma \lesssim 40.5^o$,~ 
%$106.3^o\lesssim \gamma \lesssim 140.9^o$,
$102.5^o\lesssim \gamma \lesssim 157.9^o$, 
for the central value of $R_b$ for example,
wherein the shift of $\sin(2\gamma)$ is less than 29\% of its magnitude
so that it would be measurable in this regime if the luminosity
is large enough to provide a sufficient number of events.
By measurable, we mean that a 3$\sigma$ result for its value
is not blocked by the uncertainty from penguins.
We define this regime in which $|\Delta\sin(\gamma)/\sin(\gamma)|$
is less than $0.29$ as the measurability regime.
Approximately  34\% 
of this 
regime of measurability intersects the allowed region given by 
the limits on
$\gamma$ discussed in Ref.~\cite{sbfacwkp},
$135^o\gtrsim \gamma \gtrsim 45^o$. We need to stress the following.
When the pollution in the
$\sin(2\gamma)$ is $\lesssim | \sin(2\gamma)|$, a $15-20\%$ 
accuracy  calculation of the pollution is sufficient -- when
we have as well $|\Delta(\sin(2\gamma))| < .29 |\sin(2\gamma)|$
$\sin(2\gamma)$ is directly measurable; when 
$.29 |\sin(2\gamma)|<|\Delta(\sin(2\gamma))|\lesssim | \sin(2\gamma)|$,
we measure a quantity from which we can extract $\sin(2\gamma $)
with $~20\%$ theoretical precision so that $\sin(2\gamma)$ can still be
extracted. However, when the
pollution is itself dominant  and $\sin(2\gamma)$ is $\sim 20\%$ of it,
a $20\%$ accuracy knowledge of the pollution will not
permit the extraction of $\sin(2\gamma)$. Thus, for a given precision
on the pollution, depending on the relative size of the pollution
and $\sin(2\gamma)$, one has these three regions and 
one of these is exactly that addressed as our regime of measurability,
one wherein $\sin(2\gamma)$ is measurable.
(For the $\pm1\sigma$ deviations of $R_b$, 
the measurability regimes are qualitatively
similar in size and location, with the exception that the lower regime
is absent for the $-1\sigma$ deviation case. 
So, we do not show these $\pm1\sigma$ deviation measurability regimes
separately here-- see Ref.~\cite{elsewhere} for the
corresponding plots analogous to that in Fig.~4.)
\par
The question naturally arises as to the sensitivity of our regime
of measurability to the parameters in our calculation. We now turn to this.
Since the penguin shifts plotted in Fig.~4
are determined from amplitude
ratios, they do not depend on the normalisations of the
distribution amplitudes, or the hard recoil gluon exchange coupling
in Figs. 1-3. What they do depend on are the relative strengths
of the leading and non-leading Gegenbauer coefficients~\cite{L-B}
in the distribution amplitudes, the relative strengths
of the penguin and non-penguin operators in the effective
weak Hamiltonian ( $a_2$ and the value of
$\alpha_s$ in our one-loop penguins), the quark running masses
and the light-cone fraction $x_2$ of the $\bar s$ in the $B_s$ as determined
by the Cornell model of B mesons. We have varied all of these
parameters systematically as currently allowed by the $1\sigma$
limits on them when they are taken from data or theory together
with data~\cite{elsewhere}. We find that the first
part of the regime of measurability varies from $[0^o,19^o]$
to $[0^o,58^o]$. Thus, it may even be true that some of
the allowed regime, $45^o\lesssim \gamma \lesssim 135$, overlaps this
first part of our regime of measurability. The most important
aspect of this variation is that most of it is due to 
changing the value of the running b-quark mass by just $\pm 3.5\%$
and by varying the value of $a_2$ between $.14$ and $.34$
(for reference, the variation in $\alpha_s$ is just that
generated by the $1\sigma$ variation of $\Lambda_{QCD}$ (see the following),
the variations of the non-leading Gegenbauer coefficients are the
$1\sigma$ variations as determined from their extraction from data
in Refs.~\cite{semdcy,C-Z}, the $1\sigma$ variations of the running quark
masses are as given in Ref.~\cite{masses} and the methods therein
, and the variation of $x_2$,
between $0.041$ and $0.071$, is as given by the parameter
variations allowed in Refs.~\cite{eichten:1980} -- see Ref.~\cite{elsewhere}
for further details).
If we do not vary these two measurable parameters, then
the first part of our measurability regime only varies between
$[0^o,35^o]$ to $[0^o,47^o]$, i.e., it is robust to the remaining
parameters in our calculation. In the actual precision hadron B-factory
environment, we can expect that both $m_b$ and $a_2$ will be known much better
than we know them currently from comparison with data, either 
experimental or theoretical (lattice) data. The current large
sensitivity to $m_b$ and $a_2$ of the upper boundary on the first part
of our regime of measurability is mainly academic because this
regime is already outside the preferred region of $\gamma$
and the variations we see with $m_b,a_2$ still leave most
of this first part outside the preferred region.
The second part of our regime of measurability begins at 
$102.5^o$ and ends at $157.9^o$. Upon variation of our fundamental
parameters as we described above, the beginning point varies
between $[98.2^o,105.5^0]$ and the ending point varies between
$[138^o,180^0]$ , so that the preferred region
of $\gamma$ which overlaps the second and most important
part of our regime of measurability, $[102.5^o,135^o]$
is only changed by $~^{+3.0}_{-4.3}$ degrees by the current uncertainties
in our fundamental parameters. Again, if we do not vary $a_2$ and $m_b$,
this already small effect is reduced significantly. 
We thus have a robust prediction
that $\gamma$ is measurable in the regime $[102.5^o,135^o]$.
\par
Recently, several authors~\cite{recent} have argued that
current data actually prefer the regime $36^\circ \leq \gamma \leq 97^\circ$,
although more recent theoretical analyses~\cite{rf1,hyc} would question
this conclusion. Here, we stress that, from our results in Fig. 4,
we can see that, in this new so-called preferred region,
except for the small region $86.6^\circ \leq \gamma \leq 92.7^\circ$,
the penguin shift is bounded in magnitude by a factor of 2 relative to
the actual value of $\sin(2\gamma)$ so that, as we have a $\sim 15\%$
accurate knowledge of this shift, we still may use our results
in the Appendix to radiatively correct this pollution out of
$\sin(2\gamma)$ to the $\sim 30\%$ accurate level, allowing again
a 3$\sigma$ measurement of $\sin(2\gamma)$. The use of this technique
to make fundamental tests of the SM is well-known~\cite{lepyellow-1989}.
\par
The BR's in Table 1, which remain qualitatively similar
to their values shown here under the variations of parameters just
considered, however tend to indicate that the required
luminosity would be more appropriate to hadron machines than
to an $e^+e^-$ annihilation B-factory. We note that the results in Table 1
are somewhat lower than the general range of similar results in
Refs.~\cite{others1,others2}.  For example, our highest values for the
BR just reach the lowest values in latter references.
The recent and upcoming measurements of rare B processes
can then already discriminate among various models
of these processes on the basis of decay rates alone.
To illustrate this,we note that in Ref.~\cite{pipi} we used 
our methods to compute the range 
\begin{equation}
  \begin{split}
&1.87\times10^{-6}(g_s^2(m_B^2)/g_s^2(m_B^2)|_{\Lambda^{(5)}_{QCD}=0.1\text{GeV}})^2(f_{B_d}/0.136\text{GeV})^2 \\
&\le \text{BR}(\bar B_d\rightarrow \pi^+\pi^-)\le \\
&2.63\times10^{-6}(g_s^2(m_B^2)/g_s^2(m_B^2)|_{\Lambda^{(5)}_{QCD}=0.1\text{GeV}})^2(f_{B_d}/0.136\text{GeV})^2.
  \end{split}
\label{rpipi}
\end{equation}
We note that, according to Ref.~\cite{PDG1998}, the current two-loop
value of $\Lambda^{(5)}_{QCD}$ is $237^{+26}_{-24}\text{MeV}$ and
according to Ref.~\cite{aoki:1999} the best value of $\sqrt{2}f_{B_d}$ is
now $210\pm30\text{MeV}$ so that we have the estimate
\begin{equation}
(g_s^2(m_B^2)/g_s^2(m_B^2)|_{\Lambda^{(5)}_{QCD}=0.1\text{GeV}})^2(f_B/0.136\text{GeV})^2 \cong 1.70.
\end{equation}
This means that
our result in Eq.(\ref{rpipi}) is consistent with the recent
CLEO result~\cite{cleopipi:1999} 
$\text{BR}(\bar B_d\rightarrow \pi^+\pi^-)= 4.7^{+1.8}_{-1.5}\pm0.6
\times 10^{-6}$.
Nonetheless, even if we allow the entire range which we and
the authors in Refs.~\cite{others1,others2} find for
$BR(\bar B_s\rightarrow \rho K_S)$,
we are led to suggest that the
B-factory of the SLAC-LBL/KEK type should focus its attention 
on other possible roads to $\gamma$. Others~\cite{sbfacwkp} have reached
a similar conclusion.
\par
Finally, we stress that we have found that the assumption of colour
suppression (factorisation) does not appear to work 
very well in our calculations.
This is consistent with the results in Refs.~\cite{GKP,aleks}
on the analysis of the data on the processes $B\rightarrow \Psi/J~K^{(*)}$.
We will take up the corresponding analysis with our methods
elsewhere~\cite{elsewhere}.\par

{\large\bf  Acknowledgements}
 
The author acknowledges the
kind hospitality of Prof. C. Prescott and SLAC Group A
and helpful discussions with Drs. P. Dauncey, Robert Fleischer
and Prof. L. Lanceri at various stages of this work.
\newpage

\section*{Appendix}
In this Appendix, we record for completeness the amplitude
which we have evaluated from Figs.~1-3. Specifically,
following the usual Feynman rules and the prescription
given in Ref.~\cite{L-B}, as already illustrated in Ref.~\cite{pipi},
we get the amplitude
\begin{equation}
%\begin{align}
%\begin{split}
{\cal M}({\bar{B}}_s \rightarrow \rho K_S)  ={(2\pi)^4\delta(P_{\bar B_s}
-P_\rho-P_{K_S})\over 2m_B 2E_\rho 2E_{K_S}(2\pi)^{9/2}}\left( A_Te^{-i\phi_T}e^{i\delta_T}+
                        \sum_jA_{P_j}e^{-i\phi_{P_j}}e^{i\delta_{P_j}}\right),
%                                        & + yyyy ,\\
%\end{split}
%\notag
\nonumber
\label{A1}
%\end{align}
\end{equation}
\begin{flushright}
(A.1)
\end{flushright}
where the ``would-be tree level'' contribution
to the amplitude is, from Figs.~1 and~2, given by
\begin{equation}
\begin{split}
A_Te^{-i\phi_T}e^{i\delta_T}&= \int d[y]d[w]\text{Tr}{\bigg\lgroup}{\bigg[} 
\frac{f_K\phi_K\gamma_5(\not\!P_{K_S}+m_{K_S})}{\sqrt{2}\sqrt{2}}
\frac{\left(-iG_Fa_2 V_{ub}V^*_{ud}\right)}{\sqrt{2}}m_\rho f_\rho\not\!\epsilon^*(P_\rho)(1-\gamma_5)\\
&\quad \frac{i}{\not\!P_b-\not\!q-m_b+i\epsilon}
%\left(m_\rho f_\rho\epsilon^*_\mu\right)
\left(-ig_s\lambda^c\gamma_\alpha\right)
 \frac{a_B\phi_B\gamma_5(\not\!P_B-m_B)}{\sqrt{2N_c}}
\left(-ig_s\lambda^c\gamma^\alpha\right)\\
   &+\quad \left(-ig_s\lambda^c\gamma_\alpha\right)
\frac{f_K\phi_K\gamma_5(\not\!P_{K_S}+m_{K_S})}{\sqrt{2}\sqrt{2}}
\left(-ig_s\lambda^c\gamma^\alpha\right)
 \frac{i}{\not\!P_d+\not\!q-m_d+i\epsilon}\\
&\quad \frac{\left(-iG_Fa_2 V_{ub}V^*_{ud}\right)}{\sqrt{2}}m_\rho f_\rho\not\!\epsilon^*(P_\rho)(1-\gamma_5)\frac{a_B\phi_B\gamma_5(\not\!P_B-m_B)}{\sqrt{2N_c}}{\bigg]}\frac{(-i)}{q^2}\\
&+\quad r_{ce}\int d[z]{\bigg[}\text{Tr}\{
\left(-ig_s\lambda^c\gamma^\alpha\right)
\frac{f_K\phi_K\gamma_5(\not\!P_{K_S}+m_{K_S})}{\sqrt{2}\sqrt{2}}\\
&\quad \frac{\left(-iG_Fa_2 V_{ub}V^*_{ud}\right)}{\sqrt{2}}\lambda^e\gamma^\mu(1-\gamma_5)\frac{a_B\phi_B\gamma_5(\not\!P_B-m_B)}{\sqrt{2N_c}}\}\\
&\quad {\big(}\frac{f_\rho\phi_\rho\not\!\epsilon^*(P_\rho)m_\rho}{\sqrt{2}\sqrt{2}}\left(-ig_s\lambda^c\gamma_\alpha\right)\frac{i}{\not\!P_u+\not\!q-m_u+i\epsilon}\lambda^e\gamma_\mu(1-\gamma_5)\\
&+\quad \frac{i}{-\not\!P_{\bar u}-\not\!q-m_u+i\epsilon}
\left(-ig_s\lambda^c\gamma_\alpha\right)\frac{f_\rho\phi_\rho\not\!\epsilon^*(P_\rho)m_\rho}{\sqrt{2}\sqrt{2}}\lambda^e\gamma_\mu(1-\gamma_5)
{\big)}{\bigg]}\frac{(-i)}{q^2}{\bigg\rgroup}\\
\end{split}
\nonumber
\end{equation}
\begin{flushright}
(A.2)
\end{flushright}
where contribution of Fig. 2 is (not) included for $r_{ce}=1(0)$ and
where $\frac{f_K\phi_K}{\sqrt{2}\sqrt{2}}=
\frac{f_K}{\sqrt{2}\sqrt{2}}\phi_K(y_1,y_2),
~\frac{f_\rho}{\sqrt{2}\sqrt{2}}\phi_\rho=
\frac{f_\rho}{\sqrt{2}\sqrt{2}}\phi_\rho(z_1,z_2)$ and
$\frac{a_B}{\sqrt{2N_c}}\phi_B=\frac{a_B}{\sqrt{2N_c}}\phi_B(w_1,w_2)$ are the Lepage-Brodsky distribution
amplitudes --- \\
 $\phi_K(y_1,y_2)=y_1y_2\left(1+3\beta'_K(y_2-y_1)\right)$,~ 
$\phi_\rho(z_1,z_2)=z_1z_2\left(1.348-1.74z_1+1.74z_1^2\right)$, and
$\phi_B(w_1,w_2)=\delta(w_2-x_2)$ 
are as indicated in
the text above with
$x_2= \left(m^c_s-(m^c_s+m^c_b-m_B)m^c_b/(m^c_s+m^c_b)\right)$
following the treatment of heavy mesons suggested by Ref.~\cite{L-B}
based on non-relativistic potential model considerations for example.
Here, the constituent quark masses are taken as~\cite{eichten:1980}
$m^c_s\cong 0.51$GeV and $m^c_b\cong 5.1$GeV,
so that $x_2\cong 0.0542$ when we take $m_B\cong 5.369$GeV, as we
should according to Ref.~\cite{G-K}. From Ref.~\cite{pipi} we have
$a_B=f_B/\sqrt{12}$ where $f_B$ is the B decay constant.
Here, $P_A$ is the 4-momentum
of $A$ for all $A$ and, when a parton-type occurs in 
two external wave functions
a prime is used to distinguish the two 4-momenta in an obvious way.
To be precise, let us list these internal parton momenta as follows
for Fig. 1 : 
$P_b^+=x_1m_B,~P_b^-=(m_b^2+Q_\perp^2(B))/(x_1m_B),~\vec P_{b\perp}=
\vec Q_\perp(B), 
~P_{\bar s}^+=x_2m_B,~P_{\bar s}^-=(m_s^2+Q_\perp^2(B))/(x_2m_B),
~\vec P_{\bar s\perp}=-\vec Q_\perp(B),
~Q_\perp^2(B)=x_1x_2m_B^2-x_2m_b^2-x_1m_s^2,
~P_d^+=y_1(E_K+P_{Kz}),
~P_d^-=(m_d^2+Q_\perp^2(K))/(y_1(E_K+P_{Kz})),~\vec P_{d\perp}=
\vec Q_\perp(K),
~P^{'+}_{\bar s}=y_2(E_K+P_{Kz}),P^{'-}_{\bar s}
=(m_s^2+Q_\perp^2(K))/(y_2(E_K+P_{Kz})),
~\vec P'_{\bar s\perp}=-\vec Q_\perp(K),
~Q_\perp^2(K)=y_1y_2m_K^2-y_2m_d^2-y_1m_s^2$,
where we always work to leading order in $Q_\perp^2/m_B^2$ for
all $Q_\perp$, and where we use the usual light-cone notation
with $E_K=P_K^0$ and and $P_{Kz}=P_K^3$ so that $P_K^{\pm}=P_K^0\pm P_K^3$,
etc. The $\lambda^e$ are the QCD colour matrices generating the vector
representation carried by the quarks so that $g_s$ is the QCD
coupling constant. Thus, Eq.(A.2) illustrates explicitly how the
Feynman diagrams in Figs.~1-3 are evaluated for readers unfamiliar
with the methods we used in Ref.~\cite{pipi}, for example. 
The standard trace and integration manipulations,
taking into account the definition~\cite{L-B} 
$[dx]\equiv dx_1dx_2\delta(1-x_1-x_2)$,
then lead from
Eq.(A.2) to the result
%%%%START HERE
\begin{equation}
\begin{split}
A_Te^{-i\phi_T}e^{i\delta_T}&={-iG_F\over\sqrt{2}}~{\cal F}~
%%%{\cal F}=f_Kf_\rho a_BC_F^2g_s^2\sqrt{6}
(V^*_{ud}V_{ub}/C_F)(a_2/\sqrt{2})(P_{CMS}m_B^2/Q^2)\\
        &[
%       I_{21}(m_B^3+m_K m_b m_B-2m_B^2 m_K-2m_B^2 m_b)\\
       I_{21}\left(m_B-2(m_K+m_b)+m_K m_b/m_B\right)\\
%       & +  I_{22}(m_B m_K^2-2m_B^2E_K+2m_B^2 m_K)\\
       & +  I_{22}\left(2(m_K-E_K)+ m_K^2/m_B\right)\\
%  & + (m_Bm_K^2+x_2m_B^3-2x_2m_B^2E_K-m_Bm_Km_d\\
%       & + m_d m_B^2-2m_Km_B^2x_1)(0.291/D_2)
  & +\left(x_2m_B-2(x_2E_K+x_1m_K)+m_d+m_K(m_K-m_d)/m_B\right)(0.291/D_2)\\
   &    +(-4C_1\sqrt{2}r_{ce}/(m_Ba_2))((0.291(-m_B^2+m_K^2\\
%%c     $   (2.824d0*(\hat{a}_1-\hat{a}_2)-9.12d0*(ai2-ai3)+9.12d0*(ai3-ai4))
%     & +m_\rho^2)/(2m_B^2))(1.348(\hat{a}_1-\hat{a}_2)-1.74(\hat{a}_2-\hat{a}_%3)+1.74(\hat{a}_3-\hat{a}_4))\\
     & +m_\rho^2)/(2m_B^2))(1.348\hat{a}_1-3.088\hat{a}_2+3.48\hat{a}_3-1.74\hat{a}_4)\\
%     &  -(0.166666 m_\rho^2/m_B^2)(-0.253(\hat{b}_0-\hat{b}_1)+2.505(\hat{b}_1%-\hat{b}_2)))]
     &  -(0.166666 m_\rho^2/m_B^2)(-0.253\hat{b}_0+2.758\hat{b}_1-2.505\hat{b}_2))]
\end{split}
\nonumber
\end{equation}
\begin{flushright}
(A.3)
\end{flushright}
where the various mathematical symbols are defined below.
Continuing in this way, using the entirely similar methods, we find that
the penguin graphs in Fig.~3 correspond to the contributions
to the amplitude in Eq.(A.1) given by
\begin{equation}
\begin{split}
      A_{P_1}e^{-i\phi_{P_1}}e^{i\delta_{P_1}}&={-iG_F\over\sqrt{2}}~{\cal F}~
            \{(m_BP_{CMS}/Q^2)
{\alpha_P}^{(a)}I_{31}m_B^2[\left(x_2-2(x_2E_K+x_1m_K)/m_B+m_K^2/m_B^2\right)/D_2\\
          &+\left(1-2(m_b+m_K)/m_B+m_Km_b/m_B^2\right)I_{21}/0.291+
                 (-1+2m_K/m_B+m_\rho^2/m_B^2)I_{22}/0.291]\\
%           &+ ((-2m_Bm_b-2m_Bm_K)I_{31}/0.291)I_{21}\\
%          &+   (2m_Bm_KI_{31}/0.291)I_{22})]
%%        +(appce*\alpha_P^{1}|_g+apceb*\alpha_P^1)rcep\\
&      +({\cal P}_{cea}{\alpha_P}^{(a)}|_g+{\cal P}_{ceb}{\alpha_P}^{(a)})
r_{cep}\}\\
%        TSP1=(APP-APPT)/ALP1
%%cx      APP=APP+DCMPLX(PCMS/(x_2*EK0),0)*
      A_{P_2}e^{-i\phi_{P_2}}e^{i\delta_{P_2}}&={-iG_F\over\sqrt{2}}~{\cal F}~
         (m_BP_{CMS}/Q^2)
          {\alpha_P}^{(b)}[\left(m_bm_\rho^2(x_2m_B/2-m_K)/D_2\right)I_{33} \\
%          &-(m_bm_K(-x_1^2m_B^2-x_1m_BE_K
%                 +x_1E_Km_B)/D_2)I_{32}\\
          &+\left(m_bm_Kx_1(m_\rho^2-m_K^2)+m_bm_B((1+\frac{3}{2}x_1-x_1^2)m_K^2-x_2m_B^2+(1-\frac{1}{2}x_1)x_2m_\rho^2)\right)I_{32}/D_2\\
%%              DCMPLX(.0m_bm_\rho(x_1M_K^2+x_2m_\rho^2)/RD2,0)YI33+
%%                 +DCMPLX(.0m_bm_\rho(M_K^2x_1+ M_B^2-M_K^2(1
%%                       +x_1^2)-m_\rho^2(1-x_1x_2) )/RD2,0)YI32+
%                 &+(-4m_Km_bx_1(E_K m_B-x_2m_B^2)/D_2)I_{32}\\
%%%
%                & +(-.5m_Bm_b(x_1(m_K^2-x_2E_K m_B)-x_1x_2
%                       E_K m_B+x_1x_2m_B^2)/D_2)I_{32}\\
%                & +m_bm_B\left((2-\frac{1}{2}x_1^2)m_K^2-\frac{1}{2}x_1x_2
%                       m_B^2-(2-x_1)x_2E_K m_B\right)I_{32}/D_2\\
%                 &+(-.5m_Bm_b(x_2P_\rho\cdot P_K-x_2E_\rho m_B)/D_2)I_{33}
%%              +(0m_Bm_bM_Km_\rhox_1/D_2,0)(YI33-DCMPLX(x_1,
%%                       0)YI32)+
%              +(2m_Bm_b(m_K^2-x_2 E_K m_B)/D_2)I_{32}\\
%%              +DCMPLX(.0m_bM_B^2m_\rho/.291,0)(DCMPLX(x_1,
%%                       0)YI32-YI33)(YI21-YI22)+
%%                 +DCMLX(.0m_bm_\rho/.291,0)YI22(DCMPLX(-m_\rho^2
%%                       ,0)YI33+DCMPLX(x_1m_\rho^2-x_1M_K^2,0)YI32)
%%           +DCMPLX(0m_b^2M_Bm_\rho/.291,0)YI21(YI33-DCMPLX(
%%                       x_1,0)YI32)+
%%            DCMPLX(0m_bM_BM_Km_\rho/.291,0)((YI21-YI22)YI33+
%%                       DCMPLX(x_1,0)YI32YI22)+
%%                 DCMPLX(-.0m_b^2M_Km_\rho/0.291,0)YI21YI33-
                &+\left(m_bm_B^3x_1-\frac{1}{2}x_1m_b^2m_B^2
                -4x_1m_bm_B^2E_K(1-\frac{m_b}{2m_B})+3x_1m_bm_B^2m_K(\frac{1}{2}-
                \frac{m_b}{m_B})\right)I_{32}I_{21}/0.291\\
                &+\left(3x_1m_bm_Bm_K^2-x_1m_bm_K(m_B^2+\frac{1}{2}m_K^2-
                 -\frac{1}{2}m_\rho^2)\right)I_{32}I_{22}/0.291\\
                &-m_bm_\rho^2m_B(1-m_b/(2m_B))I_{33}I_{21}/0.291\\
                &+(m_bm_Km_\rho^2/0.582)I_{33} I_{22}].
%                 +(0.5m_bm_Kx_1(m_B^2+m_K^2-m_\rho^2)/0.291)I_{32} I_{22}\\
%                & -(0.5m_bm_Kx_1m_B^2/0.291)(I_{32} I_{22}+I_{32} I_{21})
%                 +(m_b^2m_Km_Bx_1/0.291)I_{32} I_{21}\\
%                & -(m_bm_Bm_\rho^2/0.291)I_{33} I_{21}
%                 +(m_bm_B^3x_1/0.291)I_{32} I_{21}\\
%                & -(m_bm_Bm_K^2x_1/0.291)I_{32} I_{22}
%                 +(0.5m_b^2m_\rho^2/0.291)I_{33} I_{21}\\
%                 &-(0.5m_b^2x_1m_B^2/0.291)I_{32} I_{21}\\
%          &+(2m_bx_1E_K m_B/0.291)I_{32}((-2m_B+m_b)I_{21}-m_K I_{22})\\
%          &+(4m_bx_1 m_B/0.291)I_{32}(m_K^2 I_{22}+(m_K m_B/2-m_Km_b)I_{21})].
\end{split}
\nonumber
\end{equation}
\begin{flushright}
(A.4)
\end{flushright}
\par
In Eqs.(A.2)-(A.4), the following definitions have been used:
\begin{equation}
\begin{split}
{\cal F}&=f_Kf_\rho a_BC_F^2g_s^2\sqrt{3} \\
Q^2&=  ((E_K+P_{CMS})/(2m_B))(x_1m_B^2-m_b^2+m_s^2)\\ 
I_{21}&= (-0.253/m_B^2)\ell_{21}+(2.505/m_B^2)\ell_{22}\\
I_{22}&=(-0.253/m_B^2)\ell_{22}+(2.505/m_B^2)\ell_{23},\text{~for}\\
\ell_{21}&=0.403041-2.202003i\\
\ell_{22}&=-0.3794583-0.6585764i\\
\ell_{23}&=-0.5097241-0.1969674i,\\
I_{31}&= 0.0485\\
I_{32}&=(0.291/.3)(0.195517/m_B^2-0.064303i/m_B^2)\\
I_{33}&=(0.291/.3)(0.132055/m_B^2-0.0608173i/m_B^2)\\
D_2&= m_b^2+m_\rho^2-(E_\rho+P_{CMS})x_1m_B-(E_\rho-P_{CMS})(x_2m_B+
              m_b^2/m_B)-m_d^2\\
\hat{a}_1 &= -1.0-x_2 ln(x_1/x_2)-x_2\pi i\\
\hat{a}_2 &= -0.5-x_2-x_2^2\ln(x_1/x_2) -x_2^2\pi i\\
\hat{a}_3 &=  -1/3-x_2/2-x_2^2+x_2^3\ln(x_1/x_2) -x_2^3\pi i\\
\hat{a}_4 &= -0.25(x_1^4-x_2^4)-(4/3)(x_1^3+x_2^3)x_2\\
          & \quad -3x_2^2(x_1^2-x_2^2)-4x_2^3-x_2^4\ln(x_1/x_2) -x_2^4\pi i\\
\hat{b}_0 &= -\ln(x_1/x_2)-\pi i\\
       \hat{b}_1 &= \hat{a}_1 \\
       \hat{b}_2 &= \hat{a}_2 \\
      {\alpha_P}^{(a)}|_g & = \frac{\alpha_s}{2\pi}V^*_{jd}V_{jb}
    {\Bigg\lgroup} \left[\frac{1}{12}\left(\frac{1}{x_j-1}\right)+
    \frac{12}{13}\left(\frac{1}{x_j-1}\right)^2-
    \frac{1}{2}\left(\frac{1}{x_j-1}\right)^3 \right]x_j\\
     &+\quad \left[\frac{2}{3}\left(\frac{1}{x_j-1}\right)+
    \left(\frac{2}{3}\left(\frac{1}{x_j-1}\right)^2-
    \frac{5}{6}\left(\frac{1}{x_j-1}\right)^3+
    \frac{1}{2}\left(\frac{1}{x_j-1}\right)^4\right)x_j\right]\ln x_j{\Bigg\rgroup}\\
%.04008D0,0D0)*
%     c     (DCMPLX(  .00908D0*(4D0/3D0)*DLOG(RMCH/RMW),0D0)
%     c     +DCMPLX(-.0035D0*A*(4D0/3D0)*DLOG(RMUH/RMW),0D0)*
%     c      DCMPLX(DCOS(DELT),-DSIN(DELT))
%     c     +DCMPLX(.00908D0-.0035D0*A*DCOS(DELT),.0035D0*A*DSIN(DELT))*
%     c      DCMPLX(.961103D0                                ,0D0))
%      alp1g=alp1
\end{split}
\nonumber
\end{equation}
\begin{equation}
\begin{split}
  {\alpha_P}^{(a)} &= {\alpha_P}^{(a)}|_g +\frac{\alpha_{em}}{8\pi s_W^2C_F}
        {\Big\lgroup} -s_W^2\frac{16}{27}V_{cd}^*V_{cb}\ln (x_u/x_c)\\
      &+\quad V_{td}^*V_{tb}\{ -4\left(3x_t^2\ln(x_t)/(4(x_t-1)^2)+x_t/4-
          3x_t/(4(x_t-1))\right)\\
      &-\quad 2\left(5x_t/(2(x_t-1))\right)\left(1-\ln(x_t)/(x_t-1)\right)
       - 2|V_{td}|^2\{-3x_t^3\ln(x_t)/(2(x_t-1)^3)\\
      &-\quad x_t(0.25- 9/(4(x_t-1))
       -3/(2(x_t-1)^2))\}
      + (4s_W^2/3)(0.641-x_t(7/(3(x_t-1))\\
      &+\quad 13/(12(x_t-1)^2)-1/(2(x_t-1)^3))
       -x_t\ln(x_t)(1/(6(x_t-1))-35/(12(x_t-1)^2)\\
       &\quad -5/(6(x_t-1)^3)+1/(2(x_t-1)^4))
      + (2/3)^2\ln(x_u))-(x_t/2-3/(4(x_t-1))\\
     &+\quad  3(2x_t^2-x_t)\ln(x_t)/(4(x_t-1)^2)-0.75) \}{\Big\rgroup}\\
%     c     (DCMPLX(-s2w*(8d0/9d0)*.00908D0*(2D0/3D0)*DLOG(Rxu/rxc),0D0)
%     c     +DCMPLX(-.0035D0*A*0d0*(4D0/3D0)*DLOG(RMUH/RMW),0D0)*
%     c      DCMPLX(DCOS(DELT),-DSIN(DELT))
%     c     +DCMPLX(.00908D0-.0035D0*A*DCOS(DELT),.0035D0*A*DSIN(DELT))*
%     c    DCMPLX(-4d0*(3d0*rxt**2*dlog(rxt)/(4d0*(rxt-1d0)**2)+rxt/4d0-
%     c     3d0*rxt/(4d0*(rxt-1d0)))-2d0*(5d0*rxt/(2d0*(rxt-1d0)))*(1d0-
%     c      dlog(rxt)/(rxt-1d0))- 2d0*
%     ccdabs(DCMPLX(.00908D0-.0035D0*A*DCOS(DELT),.0035D0*A*DSIN(DELT)))
%     c**2     *(-3d0*rxt**3*dlog(rxt)/(2d0*(rxt-1d0)**3)-rxt*(.25d0-
%     c      9d0/(4d0*(rxt-1d0))-3d0/(2d0*(rxt-1d0)**2)))
%     c +s2w*(4d0/3d0)*( .961103D0*2d0/3d0-rxt*(7d0/(3d0*(rxt-1d0))
%     c      +13d0/(12d0*(rxt-1d0)**2)-1d0/(2d0*(rxt-1d0)**3))
%     c  -rxt*dlog(rxt)*(1d0/(6d0*(rxt-1d0))-35d0/(12d0*(rxt-1d0)**2)
%     c  -5d0/(6d0*(rxt-1d0)**3)+1d0/(2d0*(rxt-1d0)**4)) + 
%     c   (2d0/3d0)**2*dlog(rxu))-(rxt/2d0-3d0/(4d0*(rxt-1d0))+
%     c 3d0*(2d0*rxt**2-rxt)*dlog(rxt)/(4d0*(rxt-1d0)**2)-.75d0),0D0))
x_j&= m_j^2/M_W^2,~j=u,c,t,\\
{\alpha_P}^{(b)} &= \frac{-\alpha_s}{2\pi}V_{td}^*V_{tb}(-0.195)+
\frac{\alpha_{em}}{6\pi C_F}V_{td}^*V_{tb}{\Big\lgroup} 0.641
         + x_t\{1/(2(x_t-1))\\
      &+\quad 9/(4(x_t-1)^2)+3/(2(x_t-1)^3)\}
       -3x_t^3\ln(x_t)/(2(x_t-1)^4){\Big\rgroup}\\
%      ALP2=DCMPLX(-.04008D0,0D0)*DCMPLX(.00908D0-.0035D0*A*DCOS(DELT),
%     c      .0035D0*A*DSIN(DELT))*DCMPLX(-.1946488D0
%     c      ,0D0)
%      alp2=alp2+DCMPLX(1d0/(alfinv*6d0*pi*rcfcs),0D0)*
%     c  DCMPLX(.00908D0-.0035D0*A*DCOS(DELT),.0035D0*A*DSIN(DELT))*
%     c  DCMPLX(.1946488D0*2d0/3d0+rxt*(1d0/(2d0*(rxt-1d0))+
%     c  9d0/(4d0*(rxt-1d0)**2)+3d0/(2d0*(rxt-1d0)**3))
%     c  -3d0*rxt**3*dlog(rxt)/(2d0*(rxt-1d0)**4),0D0)
{\cal P}_{cea}&=( C_G/C_F) (m_B P_{CMS}/(4 x_{c2} d_{bk} d_{bp})) {\Big\lgroup}
i_{cp00}\{(-\frac{1}{2}+4x_1)m_K^2+(6-4x_1)m_\rho^2\\
     &-3x_1m_B^2-11x_1m_Km_B+
\frac{1}{2}(m_s^2-m_d^2)\}+ i_{cp10}\{5m_B^2-5m_\rho^2-6m_Bm_K\}\\
     & + i_{cp01}\{\frac{5}{2}m_B^2-\frac{3}{2}m_\rho^2+\frac{3}{2}m_K^2\}
       +(2 m_B^2/d_{br})[x_1i_{dp00}\{(x_1m_B+m_K)E_K-(2-x_2)m_B^2\\
     &  -P_\rho\cdot P_K-x_2m_Bm_K+4m_KE_\rho
       +\frac{m_\rho^2}{x_1}(1-2m_K/m_B)-m_d^2(\frac{1}{2}+\frac{2}{x_1}
(1-2m_K/m_B))+\frac{1}{2}(m_K^2+m_s^2)\}\\
     & +i_{dp01}\{2E_\rho E_K+2P_\rho\cdot P_K(x_1-m_K/(2m_B))
    +m_KE_\rho(1-5x_1-\frac{3}{2}\frac{m_K}{m_B})+2x_1m_BE\rho-\frac{E_\rho}{2m_B}m_s^2\\
   & -2m_\rho^2(x_2+\frac{m_\rho^2}{2x_1m_B^2}-\frac{E_\rho}{m_B})+
   (\frac{2m_\rho^2}{x_1m_B^2}+\frac{E_\rho}{2m_B})m_d^2\}-x_1i_{dp10}\{2m_BE_K(1-\frac{3m_K}{2m_B})+m_K^2\}\\
   &+2i_{dp11}\{P_\rho\cdot P_K(1-\frac{E_K}{m_B}-\frac{3m_K}{2m_B})+E_K E_\rho+\frac{m_K^2}{2m_B}E_\rho \}+\frac{E_\rho}{m_B}i_{dp02}\{m_K^2-m_B^2+(\frac{x_2}{x_1}-1)m_\rho^2 \}\\
& -2\frac{m_\rho^2}{x_1m_B^2}i_{dp12}P_\rho\cdot P_K ]{\Big\rgroup}\\
\end{split}
\nonumber
\end{equation}
\begin{equation}
\begin{split}
d_{bk}&= m_B^2+m_K^2-m_\rho^2\\
d_{bp}&= m_B^2-m_K^2-m_\rho^2\\
d_{br}&= m_B^2+m_\rho^2-m_K^2\\
x_{c2}&= 2Q^2/d_{bk}\\
i_{cp00}&= 0.529\\
i_{cp10}&= 0.529 (1-r_\beta)/2\\
i_{cp01}&= \delta_a/3+\delta_b/4+\delta_c/5,~\text{ where}\\
r_\beta&=0.418\\
\delta_a&= 1.348\\
\delta_b&= -1.74\\
\delta_c&= 1.74\\
\end{split}
\nonumber
\end{equation}
\begin{equation}
\begin{split}
i_{dp00}&=\delta_ai_{rdp1}(z_0)+\delta_bi_{rdp2}(z_0)+\delta_ci_{rdp3}(z_0)\\
i_{dp01}&= \delta_ai_{rdp2}(z_0)+\delta_bi_{rdp3}(z_0)+\delta_ci_{rdp4}(z_0)\\
i_{dp10}&= (\delta_ai_{rdp1}(z_0)+\delta_bi_{rdp2}(z_0)+\delta_ci_{rdp3}(z_0))(1-r_\beta)/2=i_{dp00}(1-r_\beta)/2\\
i_{dp02}&= \delta_ai_{rdp3}(z_0)+\delta_bi_{rdp4}(z_0)+\delta_ci_{rdp5}(z_0)\\
i_{dp11}&= (\delta_ai_{rdp2}(z_0)+\delta_bi_{rdp3}(z_0)+\delta_ci_{rdp4}(z_0))(1-r_\beta)/2=i_{dp01}(1-r_\beta)/2\\
i_{dp12}&= (\delta_ai_{rdp3}(z_0)+\delta_bi_{rdp4}(z_0)+\delta_ci_{rdp5}(z_0))(1-r_\beta)/2=i_{dp02}(1-r_\beta)/2
\text{, where}\\
i_{rdp1}(z)&=-1-z\ln((1-z)/z)-\pi z i\\
i_{rdp2}(z)&= -.5-z-z^2\ln((1-z)/z) -\pi z^2 i\\
i_{rdp3}(z)&= -((1-z)^3+z^3)/3
                  -3z((1-z)^2-z^2)/2
                 -3z^2-z^3\ln((1-z)/z) -\pi z^3 i\\
i_{rdp4}(z)&= -((1-z)^4-z^4)/4
                  -4z((1-z)^3+z^3)/3
                  -3z^2((1-z)^2-z^2)\\
        &\quad  -4z^3-z^4\ln((1-z)/z) -\pi z^4 i\\
i_{rdp5}(z)& = -((1-z)^5+z^5)/5
                  -5z((1-z)^4-z^4)/4
                  -10z^2((1-z)^3+z^3)/3\\
        &\quad    -5z^3((1-z)^2-z^2)
                 -5z^4-z^5\ln((1-z)/z) -\pi z^5 i 
\text{, for}\\
z_0&= m_b^2/(x_1 d_{br})\\  
{\cal P}_{ceb}&= ((1-.5 C_G/C_F)( m_B P_{CMS})/(2 Q^2))
    (-1+r_\beta) \{\delta_a i_{rdp1}(x_1)+(\delta_b-\delta_a)i_{rdp2}(x_1)\\
&\quad +(\delta_c-\delta_b) i_{rdp3}(x_1)
   -\delta_c i_{rdp4}(x_1)\},\\
\end{split}
\nonumber
\end{equation}
\begin{flushright}
(A.5)
\end{flushright}
where the kinematics is the usual two-body decay one:
$m_B=E_K+E_\rho$,~$E_K=d_{bk}/(2m_B)$, 
$P_{CMS}=\sqrt{\Delta(m_B^2,m_K^2,m_\rho^2)}/(2m_B)$ 
for $\Delta(x,y,z)=x^2+y^2+z^2-2xy-2xz-2yz$, so that the decay width itself
is given by 
\begin{equation}
\Gamma(\bar B\rightarrow \rho K_S)= |{\cal M}|^2 P_{CMS}/(8\pi m_B^2)
\nonumber
\end{equation}
\begin{flushright}
(A.6)
\end{flushright}
Here, $C_G=3,~C_F=4/3$ and we have used the values~\cite{masses} 
$m_u(1\text{GeV})\cong 5.0$MeV, 
$m_d(1\text{GeV})\cong 8.9$MeV,
$m_s(1\text{GeV})\cong .175$GeV, $m_c(m_c)\cong 1.3$GeV, $m_b(m_b)\cong 4.5$GeV,
and $m_t(m_t)\cong 176$GeV. We take $s_W^2=\sin^2\theta_W \cong 0.2315$, where
$\theta_W$ is the usual weak mixing angle; and, $\alpha_{em}$ is the QED fine
structure constsnt. We note further that we use an average value of the 
square of the momentum
transfer to the would-be
spectator in Figs. 1-3 to get $g_s^2\cong 3.72$ in ${\cal F}$ above; 
the analogous
average for the square of the momentum transfer through the penguin
yields $\alpha_s\cong .25$ in the evaluation of coefficients $\alpha_P^{(i)}$
above. Thus, in both cases, we see that the momentum transfers are large
enough that they are well into the perturbative regime where the
methods of Ref.~\cite{L-B} apply.
This completes our appendix.
\par

\end{document}